\documentclass{elsart4}%
\usepackage{amssymb}
\usepackage{amsfonts}
\usepackage{amsmath}
\usepackage{graphicx}%
\providecommand{\U}[1]{\protect\rule{.1in}{.1in}}
\begin{document}
\begin{frontmatter}
\title{Signatures of a quantum dynamical phase transition in a three-spin
system in presence of a spin environment}
\author{Gonzalo A. \'{A}lvarez,}
\author{Patricia R. Levstein,}
\author{Horacio M. Pastawski\corauthref{horacio}}
\address{Facultad de Matem\'{a}tica, Astronom{\'{\i}}a y F{\'{\i}}sica, Universidad
Nacional de C\'{o}rdoba, 5000, C\'{o}rdoba, Argentina}
\corauth[horacio]{horacio@famaf.unc.edu.ar}
\begin{abstract}
We have observed an environmentally induced quantum dynamical phase
transition in the dynamics of a two spin experimental swapping gate [J. Chem
Phys. \textbf{124}, 194507 (2006)]. There, the exchange of the coupled states
$\left\vert \uparrow,\downarrow\right\rangle $ and $\left\vert \downarrow
,\uparrow\right\rangle $ gives an oscillation with a Rabi frecuency $b/\hbar$
(the spins coupling). The interaction, $\hbar/\tau_{\mathrm{SE}}$ with a
spin-bath degrades the oscillation with a characteristic decoherence time. We
showed that the swapping regime is restricted only to $b\tau_{\mathrm{SE}%
}\gtrsim\hbar$. However, beyond a critical interaction with the environment
the swapping freezes and the system enters to a Quantum Zeno dynamical phase
where relaxation decreases as coupling with the environment increases. Here,
we solve the quantum dynamics of a two spin system coupled to a spin-bath
within a Liouville-von Neumann quantum master equation and we compare the
results with our previous work within the Keldysh formalism. Then, we extend
the model to a three interacting spin system where only one is coupled to the
environment. Beyond a critical interaction the two spins not coupled to the
environment oscillate with the bare Rabi frequency and relax more slowly. This
effect is more pronounced when the anisotropy of the system-environment
interaction goes from a purely XY to an Ising interaction form.
\end{abstract}
\begin{keyword}
Decoherence \sep swapping operation \sep Quantum Zeno Effect \sep spin dynamics \sep open
systems \sep NMR Cross Polarization
\end{keyword}
\end{frontmatter}


The control of quantum dynamics started to receive much attention because of
its relevance to several applications ranging from quantum information
processing \cite{BD2000} to nanotechnology \cite{Petta05}. Since the
environment usually acts degrading the quantum dynamics of a system
\cite{Zurek2003}, its decoupling becomes a major challenge. Many techniques
\cite{Taylor05,Petta05,Ardavan06} are developed to avoid this loss of
information that is characterized by a decoherence rate $1/\tau_{\phi}$.\emph{
}Spin systems are ideal candidates to test the procedures for QIP. Recently,
we observed an environmentally induced quantum dynamical phase transition in
the dynamics of a two spin experimental swapping gate \cite{JCP06}. There, the
exchange of the coupled states $\left\vert \uparrow,\downarrow\right\rangle $
and $\left\vert \downarrow,\uparrow\right\rangle $ gives an oscillation with a
Rabi frequency $b/\hbar$ (the spins coupling). We showed that the swapping
regime is restricted only to $b\tau_{\mathrm{SE}}\gtrsim\hbar$, where
$\hbar/\tau_{\mathrm{SE}}$ is the system-environment (SE) interaction, and
essentially within this regime $1/\tau_{\phi}\sim1/\tau_{\mathrm{SE}}$.
However, beyond a critical interaction with the environment, the swapping
freezes and the system enters to a Quantum Zeno dynamical phase where the
relaxation rate decreases as the coupling with the environment increases
$1/\tau_{\phi}\sim b^{2}\tau_{\mathrm{SE}}.$ Here, we will show how this
criticality is useful to \textquotedblleft isolate\textquotedblright\ a spin-pair.

Firstly, we solve the quantum evolution of a two spin system coupled to a
spin-bath within the Liouville-von Neumann quantum master equation
\cite{Abragam,Ernst}, and compare the result with our previous one
\cite{JCP06} using the Keldysh formalism \cite{SSC06}. Then, we extend the
model to a three interacting spin system where only one is coupled to the
environment. Beyond a critical interaction, the two spins not directly coupled
to the environment oscillate with their bare frequency and relax more slowly.
In a two spin system there is always a critical point that depends on the
anisotropy relation of the SE interaction quantified as the ratio between
Ising and XY terms. However, in the three spin system, the decoherence rate
has a\ smooth cross-over from proportional to the SE interaction to inversely
proportional to it. This cross-over approaches a critical transition as the
anisotropy of the SE interaction goes from purely XY to Ising.

Experiments in Ref. \cite{JCP06} focus on two interacting spins $1/2$ coupled
to a spin-bath as modeled by the Hamiltonian $\mathcal{H}=\mathcal{H}%
_{\mathrm{S}}+\mathcal{H}_{\mathrm{SE}}+\mathcal{H}_{\mathrm{E}}$, with%
\begin{equation}
\mathcal{H}_{\mathrm{S}}=\hbar\omega_{\mathrm{z}}\left(  I_{1}^{z}+I_{2}%
^{z}\right)  +b\left(  I_{1}^{+}I_{2}^{-}+I_{1}^{-}I_{2}^{+}\right)
/2,\label{H_system}%
\end{equation}%
\begin{equation}
\mathcal{H}_{\mathrm{SE}}=\alpha I_{2}^{z}F_{{}}^{z}+\beta\left(  I_{2}%
^{+}F_{{}}^{-}+I_{2}^{-}F_{{}}^{+}\right)  /2,\label{H_SE}%
\end{equation}%
\begin{multline}
\mathcal{H}_{\mathrm{E}}=\hbar\omega_{\mathrm{z}}%
{\textstyle\sum_{2<i}}
I_{i}^{z}\label{H_env}\\
+%
{\textstyle\sum_{2<i<j}}
b_{ij}\left[  2I_{i}^{z}I_{j}^{z}-\tfrac{1}{2}\left(  I_{i}^{+}I_{j}^{-}%
+I_{i}^{-}I_{j}^{+}\right)  \right]  ,
\end{multline}
where $\mathcal{H}_{\mathrm{S}}$ is the system Hamiltonian of the two coupled
spins, $\mathcal{H}_{\mathrm{E}}$ is the spin-bath Hamiltonian with a
truncated dipolar interaction and $\mathcal{H}_{\mathrm{SE}}$ is the SE
interaction with $F^{u}=\sum_{l}b_{2l}^{{}}I_{l}^{u},~u=x,y,z$ and $F^{\pm
}=\left(  F^{x}\pm\mathrm{i}F^{y}\right)  .$ $\mathcal{H}_{\mathrm{SE}}$ is an
Ising interaction if $\beta/\alpha=0$ and a XY, isotropic (Heisenberg) or the
truncated dipolar interaction if $\alpha/\beta=0,1,-2$ respectively. We use
the model proposed by M\"{u}ller \emph{et al. }\cite{MKBE74} to calculate the
spin dynamics in the experimental system of Ref. \cite{JCP06}. The model
assumes that only one spin interact with the spin-bath which is described in a
phenomenological way. In a quantum mechanical relaxation theory the terms
$F^{u}$ are bath operators while in the semi-classical theory
\cite{Abragam,Ernst} $F^{u}\left(  t\right)  $ represent classical stochastic
forces. As the experimental conditions justify a high temperature
approximation, the semiclassical theory coincides with a quantum treatment. By
tracing on the bath variables the random SE\ interaction Hamiltonian is
\begin{equation}
\mathcal{H}_{\mathrm{SE}}\left(  t\right)  =\alpha F_{{}}^{z}\left(  t\right)
I_{2}^{z}+\beta\left[  F_{{}}^{-}\left(  t\right)  I_{2}^{+}+F_{{}}^{+}\left(
t\right)  I_{2}^{-}\right]  .
\end{equation}
The time average of these random processes satisfy $\overline{F^{u}\left(
t\right)  }=0$, where their correlation functions are $g^{\left(  u,v\right)
}\left(  \tau\right)  =\overline{F^{u}\left(  t\right)  F^{v\ast}\left(
t+\tau\right)  }.$\emph{\ }Following the usual treatment to second order
approximation, the dynamics of the reduced density operator is given by
\cite{Abragam,Ernst}
\begin{equation}
\frac{\mathrm{d}}{\mathrm{d}t}\sigma\left(  t\right)  =-\frac{i}{\hbar
}[\mathcal{H}_{\mathrm{S}},\sigma\left(  t\right)  ]-\widehat{\widehat{\Gamma
}}\left\{  \sigma\left(  t\right)  -\sigma_{0}\right\}  ,\label{master}%
\end{equation}
where the relaxation superoperator $\widehat{\widehat{\Gamma}}$ contains the
SE interaction. It accounts for the dissipative interactions between the
reduced spin system and the spin-bath and imposes the relaxation of the
density operator towards its equilibrium value $\sigma_{0}$. We assume that
the correlation times of the fluctuations are extremely short compared with
all the relevant transition rates between eigenstates of the Hamiltonian. In
this extreme narrowing regime or fast fluctuation approximation we obtain
\begin{equation}
\widehat{\widehat{\Gamma}}\left\{  \sigma\right\}  =\frac{1}{2}\sum_{u,v}%
\xi_{u,v}\mathcal{J}^{\left(  u,v\right)  }\left(  0\right)  \left[  I_{2}%
^{u},\left[  I_{2}^{v},\sigma\right]  \right]  ,
\end{equation}
where $\mathcal{J}^{\left(  u,v\right)  }\left(  \omega\right)  =\int
_{-\infty}^{\infty}d\tau g^{\left(  u,v\right)  }\left(  \tau\right)
\exp\left\{  -\mathrm{i}\omega\tau\right\}  $ is the spectral density and
$\xi_{u,v}=\left(  \alpha\delta_{u,z}+\beta\left(  \delta_{u,x}+\delta
_{u,y}\right)  \right)  $ $\left(  \alpha\delta_{v,z}+\beta\left(
\delta_{v,x}+\delta_{v,y}\right)  \right)  $. The spatial directions are
statistically independent, i.e. $g^{\left(  u,v\right)  }\left(  \tau\right)
=0$ if $u\neq v$. Notice that the axial symmetry of $\mathcal{H}_{\mathrm{S}}$
around the $z$ axis leads to the impossibility to evaluate separately
$\mathcal{J}_{x}$ and $\mathcal{J}_{y}$, where $\mathcal{J}_{u}=\frac{1}%
{2}\mathcal{J}^{\left(  u,u\right)  }\left(  0\right)  $. Thus, they will
appear only as the averaged value $\mathcal{J}_{xy}=\left(  \mathcal{J}%
_{x}+\mathcal{J}_{y}\right)  /2.$ The superoperator now can be written as
\begin{multline}
\widehat{\widehat{\Gamma}}\left\{  \sigma\right\}  =\Gamma_{\mathrm{ZZ}%
}\left[  I_{2}^{z},\left[  I_{2}^{z},\sigma\right]  \right]  \label{gamma}\\
+\Gamma_{\mathrm{XY}}\left(  \left[  I_{2}^{x},\left[  I_{2}^{x}%
,\sigma\right]  \right]  +\left[  I_{2}^{y},\left[  I_{2}^{y},\sigma\right]
\right]  \right)  ,
\end{multline}
where $\Gamma_{\mathrm{ZZ}}=\alpha^{2}\mathcal{J}_{z}$ and $\Gamma
_{\mathrm{XY}}=\beta^{2}\mathcal{J}_{xy}$. Note that $\Gamma_{\mathrm{ZZ}}$
and $\Gamma_{\mathrm{XY}}$ contain the different sources of anisotropy. The
usual approximation considers $\mathcal{J}_{x}=\mathcal{J}_{y}=\mathcal{J}%
_{z}$ (identical correlations in all the spatial directions) and $\alpha
=\beta=1$ (isotropic interaction Hamiltonian) \cite{MKBE74}. A better
approximation considers a dipolar interaction Hamiltonian, i.e. $\alpha
=-2\beta=2$ \cite{JCP03,JCP06}. We consider the experimental initial local
polarization on site $2$, $\sigma\left(  0\right)  =\left[  \mathbf{1}%
+\beta_{\text{\textrm{B}}}\hbar\omega_{0I}I_{2}^{z}\right]  /\mathrm{Tr}%
\left\{  \mathbf{1}\right\}  $ and the spin-bath polarized, where
$\beta_{\mathrm{B}}=1/\left(  k_{\mathrm{B}}T\right)  $. As the final state
reaches the temperature of the spin-bath, $\sigma_{0}=\left[  \mathbf{1}%
+\beta_{\mathrm{B}}\hbar\omega_{0I}\left(  I_{1}^{z}+I_{2}^{z}\right)
\right]  /\mathrm{Tr}\left\{  \mathbf{1}\right\}  $. Here, $\sigma_{0}$
commutes with $\mathcal{H}_{\mathrm{S}}$, not containing coherences with
$\Delta M\geq1$.


Following the standard formalism \cite{Abragam,Ernst}, we write the
superoperator $\widehat{\widehat{\Gamma}}$ using the basis of eigenstates of
the system Hamiltonian (\ref{H_system}). After neglecting the rapidly
oscillating non-secular terms with respect to the Hamiltonian, i.e.:
$\Gamma_{\mathrm{ZZ}},\Gamma_{\mathrm{XY}}\ll b$, we solve Eq. (\ref{master})
and we calculate the magnetization of the spin $1$ obtaining an extension of
the result of Ref. \cite{MKBE74}. Our essential contribution is that we
specifically account for the anisotropy arising from the nature of SE
interaction reflecting it in $\Gamma_{\mathrm{ZZ}}=\alpha^{2}\mathcal{J}_{z}$
and $\Gamma_{\mathrm{XY}}=\beta^{2}\mathcal{J}_{xy},$%
\begin{multline}
M_{I_{1}^{z}}\left(  t\right)  =\mathrm{Tr}\left\{  I_{1}^{z}\sigma\left(
t\right)  \right\}  =M_{0}\left[  1-\frac{1}{2}e^{-\beta^{2}\mathcal{J}_{xy}%
t}\right. \label{M_isotropic}\\
\left.  -\frac{1}{2}\cos\left(  \omega_{0}t\right)  \,e^{-\left(  2\beta
^{2}\mathcal{J}_{xy}+\alpha^{2}\mathcal{J}_{z}\right)  t/2}\right]  ,
\end{multline}
where $\omega_{0}=b/\hbar$ and $M_{0}=\beta_{\mathrm{B}}\hbar\omega_{0I}/4$.

If we relax the condition $\Gamma_{\mathrm{ZZ}},~\Gamma_{\mathrm{XY}}\ll b,$
i.e., we do not neglect the non secular terms for the superoperator
$\widehat{\widehat{\Gamma}}$, the dynamics still occurs in the Liouville space
of the populations and ZQT. We obtain exactly the same solution derived in
\cite{JCP06} within the Keldysh formalism:
\begin{equation}
\frac{M_{I_{1}^{z}}\left(  t\right)  }{M_{0}}=\left(  1-a_{0}e^{-R_{0}t}%
-a_{1}\cos\left[  \left(  \omega+\mathrm{i}R_{2}\right)  t+\phi_{0}\right]
e^{-R_{1}t}\right)  , \label{PolCST}%
\end{equation}
where the real functions $\omega,$ $R_{0},R_{1}$ and $R_{2}$ as well as
$a_{0}$, $a_{1}$ and $\phi_{0}$ depend exclusively on $b,$ $1/\tau
_{\mathrm{SE}}=\Gamma_{\mathrm{ZZ}}+\Gamma_{\mathrm{XY}}$ and $p_{\mathrm{XY}%
}=\Gamma_{\mathrm{XY}}/\left(  \Gamma_{\mathrm{ZZ}}+\Gamma_{\mathrm{XY}%
}\right)  $. The complete analytical expression is given in Ref. \cite{JCP06}.
There, we showed that if the SE interaction is \emph{anisotropic}
($\Gamma_{\mathrm{XY}}\neq\Gamma_{\mathrm{ZZ}}$), the functional dependence of
$\omega\ $on $\tau_{\mathrm{SE}}$ and $b$ yields a critical value for their
product, $b\tau_{\mathrm{SE}}/\hbar=k_{p_{\mathrm{XY}}},$ where the dynamical
regime changes. We called this phenomenon a quantum dynamical phase transition
\cite{JCP06,SSC06} ensured by the condition $\omega R_{2}\equiv0$. This is
evidenced by the functional change (non-analyticity) of the dependence of the
observables (e.g. the swapping frequency $\omega$) on the control parameter
$b\tau_{\mathrm{SE}}/\hbar$. This non-analyticity is enabled by taking the
thermodynamic limit of an infinite number of spins.\cite{sachdev}. One
identifies two parametric regimes: 1- The \textit{swapping phase,} which is a
form of sub-damped dynamics, when $b\tau_{\mathrm{SE}}/k_{p_{\mathrm{XY}}%
}>\hbar$ ($R_{2}=0$ in Eq. (\ref{PolCST})). 2- A \textit{Zeno phase}, with an
over-damped dynamics for $b\tau_{\mathrm{SE}}/k_{p_{\mathrm{XY}}}<\hbar$
arising on the strong coupling with the environment\ (\emph{zero frequency},
i.e. $\omega=0$, in Eq. (\ref{PolCST})). In the neighborhood of the critical
point the swapping frequency takes the form:%
\begin{equation}
\omega=\left\{
\begin{array}
[c]{cc}%
a_{p_{\mathrm{XY}}}^{{}}\sqrt{\left(  b/\hbar\right)  _{{}}^{2}%
-k_{p_{\mathrm{XY}}}^{2}/\tau_{\mathrm{SE}}^{2}} & b\tau_{\mathrm{SE}%
}/k_{p_{\mathrm{XY}}}>\hbar\\
0 & b\tau_{\mathrm{SE}}/k_{p_{\mathrm{XY}}}\leq\hbar
\end{array}
\right.  .
\end{equation}
The parameters $a_{p_{\mathrm{XY}}}$ and $k_{p_{\mathrm{XY}}}$ only depend on
$p_{\mathrm{XY}}$ which is determined by the anisotropy of the interaction
Hamiltonian.

\begin{figure}
[tbh]
\begin{center}
\includegraphics[
height=3.8795in,
width=2.9992in
]%
{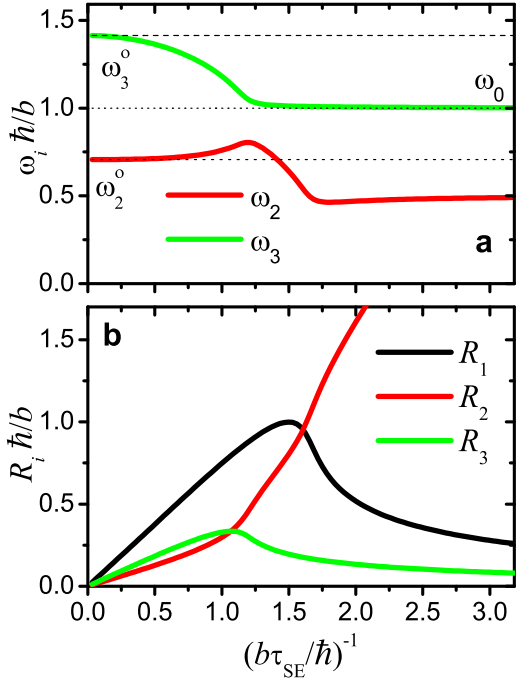}%
\caption{(color online) a) Frequencies involved in the time evolution of the
polarization in the $3$-spin system as a function of $\left(  b\tau
_{\mathrm{SE}}/\hbar\right)  ^{-1}$. Dashed lines represent the isolated
system. Dot line corresponds to two spins decoupled from the environment. b)
Different relaxation rates of the polarization.}%
\label{Fif_w_R}%
\end{center}
\end{figure}

\begin{figure}
[tb]
\begin{center}
\includegraphics[
trim=0.000000in 0.000000in 0.000000in -0.022513in,
height=3.5561in,
width=3.109in
]%
{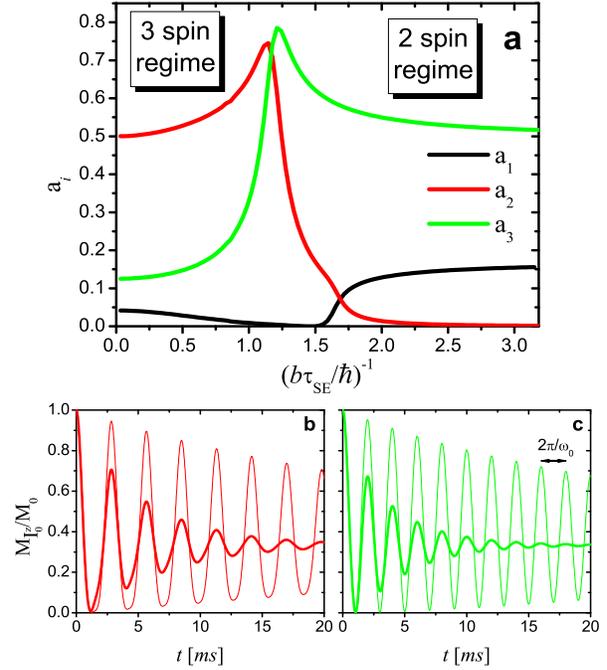}%
\caption{(color online) a) Coefficients (weights) of the different terms of
Eq. (\ref{M3spins}). At the critical region there is a switch between \ the
$2$-spin and the $3$-spin regime. b) and c) Temporal evolutions of the
polarization in the $2$-spin and $3$-spin regimes respectively for different
$\tau_{\mathrm{SE}}.$ In b) $b/\hbar=2\pi\times1\operatorname{kHz}$ and
$\tau_{\mathrm{SE}}=1.43\operatorname{ms}$ for the thick line and
$\tau_{\mathrm{SE}}=10\operatorname{ms}$ for the thin line. In c)
$b/\hbar=2\pi\times1\operatorname{kHz}$ and $\tau_{\mathrm{SE}}%
=0.1\operatorname{ms}$ for the thick line and $\tau_{\mathrm{SE}%
}=0.01\operatorname{ms}$ for the thin line.}%
\label{Fig_as_pol}%
\end{center}
\end{figure}

\textbf{Three spin system}. The system Hamiltonian is
\begin{multline}
\mathcal{H}_{\mathrm{S}}=\hbar\omega_{\mathrm{L}}\left(  I_{0}^{z}+I_{1}%
^{z}+I_{2}^{z}\right) \\
+b\left(  I_{0}^{+}I_{1}^{-}+I_{0}^{-}I_{1}^{+}\right)  /2+b\left(  I_{1}%
^{+}I_{2}^{-}+I_{1}^{-}I_{2}^{+}\right)  /2,
\end{multline}
and the environment and SE Hamiltonian remains as before. Also, the
environment is coupled to only one spin of the system. We solve Eq.
(\ref{master}) as before, without neglecting non-secular terms of the
relaxation superoperator. Considering the initial condition $\sigma\left(
0\right)  =\left(  \mathbf{1}+\beta\hbar\omega_{0I}I_{0}^{z}\right)
/\mathrm{Tr}\left\{  \mathbf{1}\right\}  $ and a polarized spin-bath, the
magnetization at site $0$ is%
\begin{multline}
M_{I_{0}^{z}}\left(  t\right)  =\mathrm{Tr}\left\{  I_{0}^{z}\sigma\left(
t\right)  \right\}  =M_{0}\left[  1-a_{0}e^{-R_{0}t}-a_{1}e^{-R_{1}t}\right.
\label{M3spins}\\
\left.  +a_{2}\sin\left(  \omega_{2}t+\phi_{2}\right)  e^{-R_{2}t}+a_{3}%
\sin\left(  \omega_{3}t+\phi_{3}\right)  e^{-R_{3}t}\right]  .
\end{multline}
The coefficients $a_{i},$ $R_{i}$, $\omega_{i}$ and $\phi_{i}$ are real and
they are functions of $b,$ $1/\tau_{\mathrm{SE}}\ $and $p_{\mathrm{XY}}.$ If
$p_{\mathrm{XY}}\neq0$ the final state has all the spins polarized because a
net transfer of magnetization from the spin-bath is possible. However, for an
Ising SE interaction, $p_{\mathrm{XY}}=0,$ we obtain that $R_{0}=0$ and
$1-a_{0}=1/3$ (the asymptotic polarization) because the final state is the
quasi-equilibrium of the $3$-spin system \cite{JCP03}. In Fig. \ref{Fif_w_R}
we show the frequencies $\omega_{2}$ and $\omega_{3}$ and the different
relaxation rates as a function of $\left(  b\tau_{\mathrm{SE}}/\hbar\right)
^{-1}$ when the SE interaction is Ising $(p_{\mathrm{XY}}=0)$. Two changes,
resembling the critical behavior of two spin systems are observed. The same
phenomenon occurs in Figure \ref{Fig_as_pol} a) where the coefficients $a_{i}$
are shown. The polarization evolution of an isolated $3$-spin system is
$M_{I_{0}^{z}}\left(  t\right)  =\frac{M_{0}}{8}\left[  3+4\cos\left(
\omega_{2}^{o}t\right)  +\cos\left(  \omega_{3}^{o}t\right)  \right]  $ where
$\omega_{2}^{o}=\left(  \sqrt{2}/4\right)  b/\hbar$ and $\omega_{3}%
^{o}=\left(  \sqrt{2}/2\right)  b/\hbar$ are the natural frequencies. When
$\left(  b\tau_{\mathrm{SE}}/\hbar\right)  ^{-1}<<1,$ we observe that
$\omega_{2}\rightarrow\omega_{2}^{o},$ $\omega_{3}\rightarrow\omega_{3}^{o},$
$a_{1}\rightarrow1/3-3/8,$ $a_{2}\rightarrow1/2$ and $a_{3}\rightarrow1/8$ as
expected for an isolated $3$-spin dynamics. The dependence of $\omega_{3}$ as
a function of $\left(  b\tau_{\mathrm{SE}}/\hbar\right)  ^{-1}$ is similar to
that of the swapping frequency in Ref. \cite{JCP06}. However, instead of
becoming zero when the SE interaction increases, it suddenly stabilizes at
$\omega_{0}=b/\hbar$, the bare $2$-spin Rabi frequency. At the same point
$\omega_{2}$, $R_{2}$ and $R_{3}$ also have a sudden change. While $R_{2}$ and
$R_{3}$ initially grow as $\left(  b\tau_{\mathrm{SE}}/\hbar\right)  ^{-1}$,
there $R_{2}$ increases the growing speed while $R_{3}$ begins to decay as in
the Zeno phase of Ref. \cite{JCP06}. Moreover, the coefficients $a_{2}$ and
$a_{3}$ switch between them, $a_{2}$ suddenly goes down and $a_{3}$ goes up.
These coefficients are the weights of the different frequency contributions in
the time evolution. The changes on the decoherence rates and on the weight
coefficients of the different oscillatory terms beyond the critical
interaction (region) lead the system to oscillate with the bare Rabi frequency
of two spins decoupled from the environment. If we keep increasing the control
parameter $\left(  b\tau_{\mathrm{SE}}/\hbar\right)  ^{-1},$ this effect is
enhanced by the next transition. After the second transition, $R_{1}$ begins
to decrease as in the Zeno phase behavior of Ref. \cite{JCP06}. As the term of
Eq. \ref{M3spins} that relaxes with $R_{1}$ leads the system to the $3$-spin
quasi-equilibrium, when $R_{1}$ goes down, this final state is approached much
slower. The effect is to avoid the interaction between the $2$-spins not
coupled to the environment and $I_{2}$. After the second transition, the
coefficient $a_{1}$ goes abruptly to zero leading to a more pronounced
\textquotedblleft isolation\textquotedblright\ of the $2$-spins. Thus, two
dynamical regimes are observed: One characterized by the $3$-spin dynamics for
$\left(  b\tau_{\mathrm{SE}}/\hbar\right)  ^{-1}\lesssim1$ and a second one,
for $\left(  b\tau_{\mathrm{SE}}/\hbar\right)  ^{-1}\gtrsim1,$ which has a
$2$-spin behavior.%

Fig. \ref{Fig_as_pol} b) and c) show the temporal evolution of the
magnetization of Eq. (\ref{M3spins}) on the $3$-spins and the $2$-spin regimes
respectively. While in Fig. \ref{Fig_as_pol} b) the two frequency
contributions are evident, in Fig. \ref{Fig_as_pol} c) only the bare Rabi
frequency is manifested. In each graph we show two curves with different SE
interactions. In Fig. \ref{Fig_as_pol} b), we show that increasing the SE
interaction the decoherence rates increase. However, in the $2$-spin regime
(Fig. \ref{Fig_as_pol} c)) when the SE interaction is increased, the
decoherence rate decreases leading to a better \textquotedblleft
isolation\textquotedblright. It is important to take into account that while
the relaxation rates go to zero smoothly the swapping frequency acquire the
bare value near the critical point. Another fact to remark is that this effect
is more pronounced when the anisotropy of the SE interaction is close to a
pure Ising SE interaction while an increase in the XY nature leads to a
further smoothing of the transition. The reason is that, when $p_{\mathrm{XY}%
}\neq0$, there is a net transfer of magnetization to the system which is
redistributed between the three spins, this redistribution begins to be slower
at the second transition when $R_{3}$ goes down. In contrast, for a purely
Ising interaction, there is no net polarization transfer and a purely
decoherent process at site $3$ freezes its dynamics but its fast energy
fluctuations prevent the interaction with the other spins.

In summary, we found an analytical expression for the two spin dynamics plus a
spin-bath of the experimental swapping gate \cite{JCP06}, showing that
standard density matrix formalism leads to a quantum dynamical phase
transition as does the Keldysh formalism \cite{JCP06,SSC06}. Here, we extended
the model to a $3$-spin system and showed that beyond a critical region the
two spins become almost decoupled from the environment oscillating with the
bare Rabi frequency and relaxing more slowly. While in the two spin swapping
gate the dynamical transition is critical, when we extend the system to
$3$-spin the criticality is smoothed out. However, enough abruptness remains
to give the possibility to use it to \textquotedblleft
isolate\textquotedblright\ a two spin system with a finite system-environment interaction.

Support from Fundaci\'{o}n Antorchas, CONICET, ANPCYT and SECYT-UNC is acknowledged.




\begin{thebibliography}{99}                                                                                               %






\bibitem {BD2000}C. H. Bennett and D. P. DiVincenzo, Nature \textbf{404}, 247 (2000).

\bibitem {Petta05}J. R. Petta, et al., Science \textbf{309}, 2180 (2005).

\bibitem {Zurek2003}W. H. Zurek, Rev. Mod. Phys\textit{.} \textbf{75}, 715 (2003).

\bibitem {Taylor05}J. M. Taylor, \emph{et al.}, Nature Phys. \textbf{1}, 177 (2005).

\bibitem {Ardavan06}J. J. L. Morton, \emph{et al.}, Nature Phys. \textbf{2},
20 (2006).

\bibitem {JCP06}G. A. \'{A}lvarez, E. P. Danieli, P. R. Levstein and H. M.
Pastawski, J. Chem Phys. \textbf{124}, 194507 (2006).

\bibitem {Abragam}A. Abragam, \textit{The Principles of Nuclear Magnetism}
(Clarendon Press, Oxford, 1961).

\bibitem {Ernst}R.R. Ernst, G. Bodenhausen, and A. Wokaun, \textit{Principles
of Nuclear Magnetic Resonance in One and Two Dimensions} (Oxford University
Press, Oxford, 1987).

\bibitem {SSC06}E. P. Danieli, G. A. \'{A}lvarez, P. R. Levstein and H. M.
Pastawski, Solid State Comm. \textbf{141}, 422 (2007).

\bibitem {MKBE74}L. M\"{u}ller, A. Kumar, T. Baumann and R. R. Ernst, Phys.
Rev. Lett\textit{.} \textbf{32}, 1402 (1974).

\bibitem {JCP03}A. K. Chattah \textit{et al.}, J. Chem. Phys\textit{.}
\textbf{119}, 7943 (2003).

\bibitem {sachdev}S. Sachdev, \textit{Quantum Phase Transitions} (Cambridge U.
P., 2001).
\end{thebibliography}
\end{document}